\documentclass[oldversion]{aa}  
\usepackage[pdftex]{graphicx}
\usepackage{txfonts}
\usepackage{natbib}

\newcommand{\kms}{km\,s$^{-1}$}
\newcommand{\degre}{$^{\circ}$}
\newcommand{\dix}[1]{$\times\mathrm{10}^{\mathrm{#1}}$}
\newcommand{\msol}{M$_{\odot}$}
\newcommand{\hms}[3]{#1$^{\mathrm{h}}$#2$^{\mathrm{m}}$#3$^{\mathrm{s}}$}
\newcommand{\dms}[3]{#1$^{\circ}$#2\arcmin#3\arcsec}
\newcommand{\mum}{\,$\mu$m}
\newcommand{\Ha}{H$_{\alpha}$}
\newcommand{\mpc}{M$_{\odot}$\,pc$^{-2}$}
\newcommand{\Xunit}{cm$^{-2}$\,(K\,\kms )$^{-1}$}
\def \H2{H$_{2}$}

\def \ratioo{N({\rm H}_2) / I_{\rm CO}}
\def \ratiob{N({\rm H}_2) / I_{\rm CO(2-1)}}
\def \ratioa{N({\rm H}_2) / I_{\rm CO(1-0)}}

\begin{document}
   \title{Particularly Efficient Star Formation in M\,33}

   \subtitle{}

  \author{E. Gardan\inst{1};
         J. Braine\inst{1};
         K.F. Schuster\inst{2};
         N. Brouillet\inst{1};
         A. Sievers\inst{2}}
  \offprints{gardan@obs.u-bordeaux1.fr}

   \institute{Universit\'e Bordeaux 1 ; CNRS ; Laboratoire d'Astrophysique, Observatoire de Bordeaux, OASU ; UMR 5804, Floirac, F-33270
         \and
              IRAM, 300 Rue de la piscine, F-38406 S$^{t}$ Martin d'H\`eres, France}
   \date{submitted 04/25/07, Accepted 07/12/07}

\abstract{The Star Formation (SF) rate in galaxies is an important parameter at all redshifts and evolutionary stages of galaxies. In order to understand the increased SF rates in intermediate redshift galaxies one possibility is to study star formation in local galaxies with properties frequently found at this earlier epoch like low metallicity and small size. We present sensitive observations of the molecular gas in M\,33, a small Local Group spiral at a distance of 840\,kpc which shares many of the characteristics of the intermediate redshift galaxies. The observations were carried out in the CO(2--1) line with the HERA heterodyne array on the IRAM 30\,m telescope. A 11\arcmin$\times$22\arcmin\ region in the northern part of M\,33 was observed, reaching a detection threshold of a few 10$^{3}$\,\msol. The correlation in this field between the CO emission and tracers of SF (8\mum, 24\mum, \Ha, FUV) is excellent and CO is detected very far North, showing that molecular gas forms far out in the disk even in a small spiral with a subsolar metallicity. One major molecular cloud was discovered in an interarm region with no HI peak and little if any signs of SF -- without a complete survey this cloud would never have been found. The radial dependence of the CO emission has a scale length similar to the dust emission, less extended than the \Ha\ or FUV. If, however, the $\ratioo$ ratio varies inversely with metallicity, then the scale length of the H$_2$ becomes similar to that of the \Ha\ or FUV. Comparing the SF rate to the H$_2$ mass shows that M\,33, like the intermediate redshift galaxies it resembles, has a significantly higher SF efficiency than large local universe spirals. The data presented here also provide an ideal test for theories of molecular cloud formation and cover a new region in parameter space, where $\Sigma_{\mathrm{stars}} < \Sigma_{\mathrm{gas}}$. We find that a simple pressure-based prescription for estimating the molecular to atomic gas fraction does not perform well for M\,33, at least in the outer parts.  On the other hand, we show that the molecular gas fraction is influenced by (i) the total Hydrogen column density, dominated in M\,33 by the HI, and (ii) the galactocentric distance.}


 \maketitle
%

\section{Introduction}

\begin{figure*}[!ht]
   \centering
   \includegraphics[angle=270.,width=500pt]{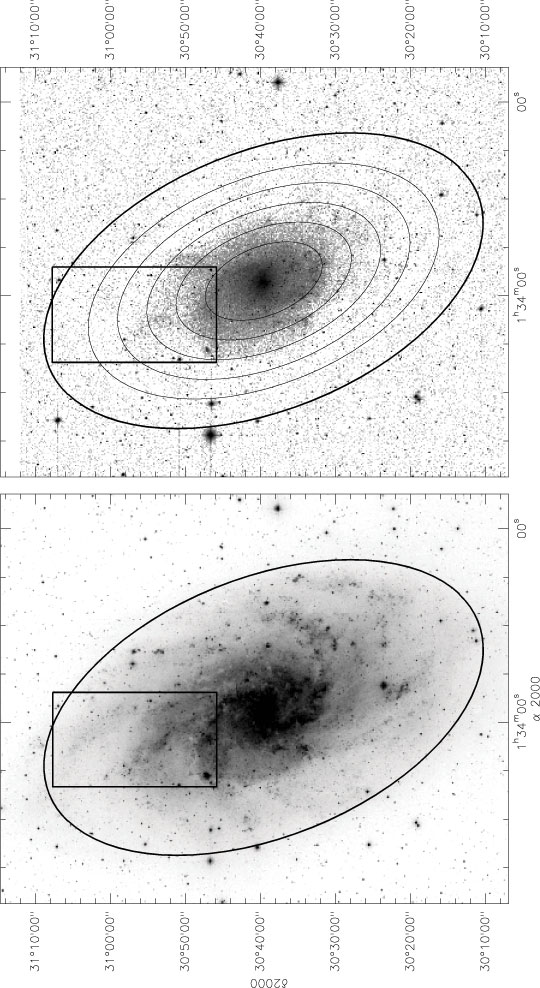}
   \caption{DSS Blue image (left) and 2MASS Ks band (right) images of M\,33. The Ks band image is smoothed with 6\arcsec\ width gaussian to bring out low-brightness features. The young stellar population (blue) extends further than the old one (K band).  The bold ellipse represents the R$_{25}$ radius (where the B band surface brightness falls below 25\,mag\,arcsec$^{-2}$). The other ellipses have semi-axes of 2, 3, 4, 5 and 6\,kpc. The box shows the area we observed with HERA. Our map covers the bright arm in the south and reaches the edge of the optical disk (R$_{25}$). In the text we refer to the "main", "middle", and "outer" arms in our map. These correspond to the features seen within the box on the Blue band image near the bottom, middle, and top respectively.}
   \label{CO_box}%
\end{figure*}

The Star Formation Rate (SFR) increases by at least a factor 10 in galaxies up to a redshift of about $z = 0.5 - 1$ \cite[e.g.][]{Madau96,Heavens04}. Since stars form from H$_2$, and not directly from HI (with the possible exception of the so-called Pop. III, or first generation stars), this suggests that either large amounts of molecular hydrogen were available or that for some reason the efficiency of star formation (SFE, defined as the SFR per unit H$_2$) was particularly high back then. In fact, the SFRs proposed are so much higher than the SFR today that both  possibilities may be required. While more galaxies existed, the bulk of the change is in the individual objects rather than having 15 -- 20 times more galaxies at $z = 0.5 - 1$. Because at least 10\% of the baryons in galaxies are thought to be in  neutral gas (and more than 10\% in many cases), an SFR a factor 15 -- 20 higher must result at least partially from a higher SFE. While there may not be a 100\% consensus, it is widely believed that the fraction of molecular gas (with respect to atomic) decreases going to later types and smaller objects \citep{Young91}. Moderate to high redshift galaxies are typically smaller and more gas-rich than today's spirals and most likely have a slightly subsolar metallicity. They thus resemble todays small spirals such as M\,33 or NGC\,2403, or the even smaller NGC\,6822, and could be expected to have a low H$_2$/HI mass ratio. If so, this would make the SFE in the objects even more extreme. How does the relationship between HI, H$_2$, and star formation depend on galaxy type, and thus redshift? M\,33 is an excellent object for this sort of study as it is a nearby small blue spiral with subsolar metallicity -- properties that seem appropriate for comparison with intermediate redshift objects. The recent GALEX images of M\,33 \citep{Thilker05} are much more extended than the Near-InfraRed (NIR) images by the 2MASS survey or the Spitzer satellite at 3.6$\mu$m, showing that we see the outer disk of M\,33 as it is growing and in the absence of strong gravitational perturbations.

In this work we present CO(2-1) observations of an 11\arcmin$\times$22\arcmin\ region in the Northern part of the Local Group spiral M~33 (NGC~598, Triangulum). M\,33 is an SA(s)cd (late type spiral with no bar and no ring) galaxy with a rotation speed of about 100\,\kms and a luminosity about 10 times (2.5 magnitudes) lower than M31 (or the Milky Way). We take a distance of 840 kpc \citep{Galleti04} for all calculations but distances vary from 800 kpc \citep{Lee02} to 900 kpc \citep{Kim02}.  The 11\arcsec\ resolution of our CO observations then corresponds to a linear size of 45 pc, sufficient to resolve individual Giant Molecular Clouds (GMCs). Similar observations have already been carried out by \citet{Engargiola03} with the BIMA array in the CO(1--0) line at 13\arcsec\ resolution.  However, the two data sets are quite complementary because their observations cover both sides of the disk whereas our observations are much more sensitive but cover a smaller region, starting near their northernmost detections and extending to the optical radius (roughly R$_{25}$). 

Like most small spiral galaxies, M\,33 has a metallicity that is subsolar by a factor 2 or 3 \citep[averaged over the disk, e.g.][]{Garnett97}. We take the solar metallicity, or more precisely Oxygen abundance, to be $12+log(O/H) = 8.67$ \citep{Asplund06}.  As we are interested in the relationship between the atomic and molecular gas and star formation, it is important to take into account the radial metallicity gradient.  When the project was initiated and the observations carried out, there was a general consensus on a metallicity gradient of roughly 0.1 dex kpc$^{-1}$ \citep{Vilchez88,Garnett97,Magrini04,Beaulieu06}, such that the outer parts of M\,33 had a metallicity of 0.1 solar, below that of the Small Magellanic Cloud.  Recently, \citet{Tiede04} found a shallower gradient, $-0.06 \pm 0.01$ dex kpc$^{-1}$, and \citet{Crockett06} a still shallower gradient, $\la -0.03$ dex kpc$^{-1}$, such that the variation could simply be another factor 2.  Either way, is molecular gas detectable far out in small galaxies such as M\,33? Are molecular clouds found only in the HI arms, which are well-correlated with the GALEX UV-bright regions?  Recently \citet{Braine04} showed that significant quantities of molecular gas were detectable far out in at least one spiral disk but that long integration times were required to detect it.  The outskirts of spirals share the subsolar metallicities and low mass surface densities of small and/or intermediate redshift spirals but not the level of star formation.  Pressure is often considered to be a major driver of the conversion of atomic gas into molecular gas, whether due to compression in spiral arms or to the ambient pressure \citep{Elmegreen94,Wong02,Blitz06}.  The observations presented here provide one of the best datasets to test these theories.

\begin{table}[!h]
    \caption{Adopted parameters for M\,33.}

    \label{carac}
      
    \begin{tabular}{ll}
     \hline
     \hline
     $\alpha$ (J2000)$^{(a)}$  &  \hms{01}{33}{51.02} \\
     $\delta$ (J2000)$^{(a)}$  &  \dms{+30}{39}{36.7} \\
     $\alpha_{0}$ $^{(b)}$     &  \hms{01}{34}{13.0}  \\
     $\delta_{0}$ $^{(b)}$     &  \dms{+30}{58}{00}   \\
     Distance$^{(c)}$          &  840 kpc             \\
     Optical radius R$_{25}$ $^{(d)}$&  30.8\arcmin   \\ 
     Inclination$^{(d)}$       &  56\degre            \\
     Position angle$^{(d)}$    &  22.5\degre          \\
     \hline
     \end{tabular}
          
     \begin{list}{}{}
      \item[(a)] center of M\,33 \citep{Cotton99}
      \item[(b)] Reference position of the map.
      \item[(c)] \citet{Galleti04}
      \item[(d)] LEDA \citep{Paturel03}, but Ks diameter is significantly smaller (Fig.~\ref{CO_box} and 
      \citet{Jarrett03}), such that the B/Ks size ratio is greater than for e.g. M\,31.
     \end{list}

\end{table}

\section{Observations and data reduction}

We observed M\,33 with the IRAM 30m telescope in June 2005 using HERA \citep{Schuster04}, a 9 receiver dual polarization array. The On-The-Fly mode was used to observe the $^{12}$CO(2-1) transition at 230.53799 GHz. We applied a drift speed of 2\arcsec/s with a dump time of 3\,s resulting in a sampling of 6\arcsec\ in the scanning direction. The scan length was 660\arcsec\ with calibrations at the beginning and the end of each scan. The OFF position was taken at $\Delta \alpha$=+900\arcsec\ with respect to the reference position $\alpha_{0}$,$\delta_{0}$ (see Tab.~\ref{carac}). 

The area was scanned both in right ascension and declination. The resulting map size is 700\arcsec$\times$1300\arcsec\ (see Fig.~\ref{CO_box}) for a total observing time of 96\,h. The typical system temperature was between 400\,K and 600\,K and opacity around 0.2. We used the WILMA backend which provides a spectral resolution of 2\,MHz, i.e 2.6\,\kms\ at this frequency. The total bandwidth of the backend WILMA is 930\,MHz, and covers the range of velocity from -500 to 500\,\kms. The rms noise level in the final data cube is 20-30\,mK in the bright southern part and $\sim$15\,mK in the North. The beam efficiency at 230.5\,GHz is $\Omega_{\mathrm{eff}}$=0.52 and the forward efficiency F$_{\mathrm{eff}}$=0.95 \footnote{see http://www.iram.es/}. Maps hereafter are presented in main beam temperature (1.8 times higher than in T$_a^*$).

All data were reduced with the CLASS90 program, a new part of the GILDAS software package. The complete dataset comprises 330,000 spectra -- such a huge dataset cannot be reduced by hand. After inspecting many spectra by hand, covering different observing periods and conditions, we wrote script in the CLASS language to automatically reduce the data and produce a datacube. First of all, spectra from the two bad receivers of the instrument are completely rejected from the dataset. Most of the spectra show a ripple due to a standing wave between the telescope secondary and the receiver. The ripple has a well-known frequency \citep{Schuster04} so frequencies between 0.143MHz$^{-1}$ and 0.155MHz$^{-1}$ are removed on all spectra using the FFT routine of CLASS90. Then a baseline of order 5, determined between -400 and -100\,\kms, is substracted from each spectrum.  A velocity dependent window, based on the rotation curve of M\,33, is excluded from this calculation.  Lines in M\,33 are very narrow, rarely greater than 10\,\kms\ and never as wide as 30\,\kms\ (and only reaching 15\,\kms\ in the NGC 604 region), so a 5th order polynomial cannot create spurious lines. The velocity range of the window where signal is expected varies as follows with the declination : spectra below $\Delta \delta\leq$-10\arcmin\ have a range between -220 and -260\,\kms; spectra with -10\arcmin$<\Delta \delta\leq$-5\arcmin\ have range of -250 to -270\,\kms; and spectra above $\Delta \delta>$-5\arcmin\ have a range between -255 and -280\,\kms; the region around NGC\,604 (\hms{01}{34}{33.2}, \dms{30}{47}{5.6}) has a range between -208 and -250\,\kms.  The observing time of individual spectra is very short (3\,s) so the lines are not apparent when the baselines are subtracted. after baseline fitting, the noise level calculated outside of the line range was compared with the noise level expected based on the system temperature.  If the noise level was significantly higher (30\%) than expected, the spectra were eliminated on the basis that something was wrong (very poor baseline fit or some instrumental oddity). Only 1300 spectra were eliminated in this way, such that the database of spectra could only be improved.

The pixel size of the output cube was chosen to be 4\arcsec, which is appropriate for a beam size of about 11\arcsec. For each spatial pixel in the output cube, the spectra around that position were added with a weight depending on the distance from the pixel center (gaussian weighting) and on the noise in the spectrum, the weight varying inversely with the square of the noise level. Only channels from -100 to -400\,\kms were kept as the lines in this part of the galaxy are around -250\,\kms and narrow. Several cubes were made, weighting the spectra with gaussians of half-power width of 7, 10, 13, and 28 arcseconds, yielding final spatial resolutions of 13, 15, 17, and 30 arcseconds respectively. Each channel was treated independently such that this process has no effect on the spectral resolution. The gridding program supplied with CLASS was not used.

   \begin{figure*}[!ht]
   \centering
   \includegraphics[width=400pt]{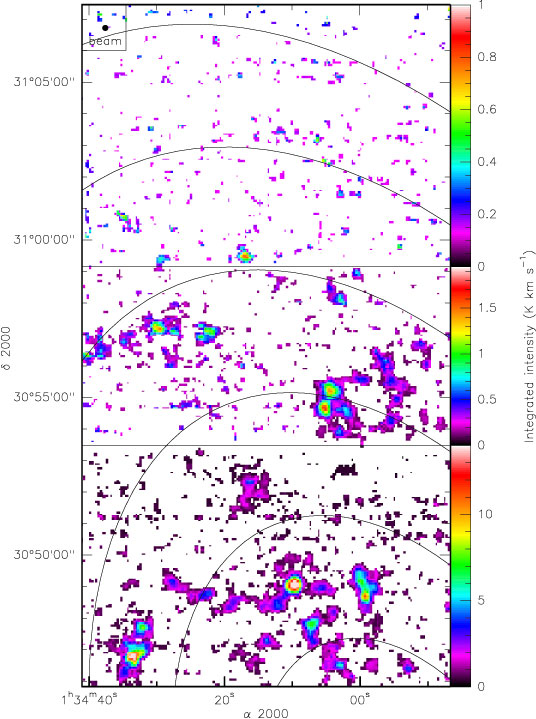}
   \caption{CO integrated intensity map obtained from the cube. The three areas correspond to the three different noise levels (30, 20 and 12\,mK from bottom to top) and cloud velocities. The velocities integration range is different for each part in order to take into account the rotation curve. Velocity ranges are : -210 to -260\,\kms, -250 to -280\,\kms, and -255 to -282\,\kms (from bottom to top). We sum channels above 3$\sigma$ to take into account of the pixel to pixel noise variation. Ellipses have semi-axes of 2, 3 .. 7\,kpc.}
   \label{I_CO}
   \end{figure*}

   \begin{figure*}[!ht]
   \centering
   \includegraphics[width=500pt]{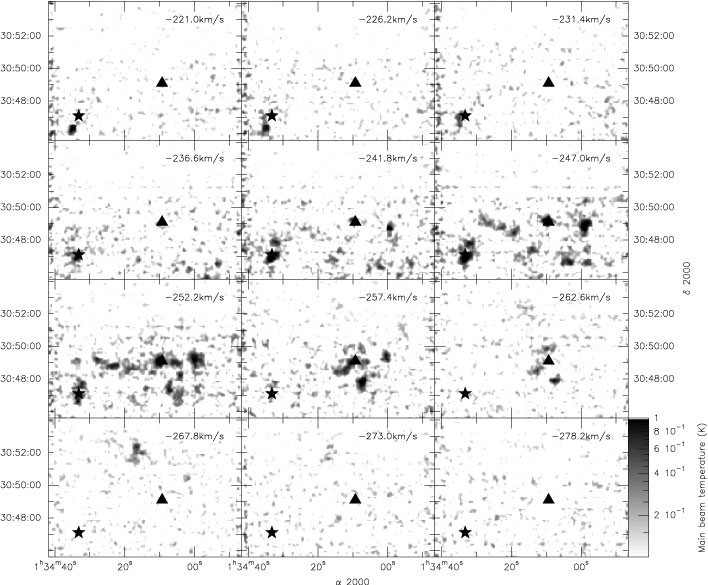}
   \caption{CO(2-1) channel maps of the southern part of the map ($\delta<$\dms{30}{54}{00}). Each panel is the sum of two adjacent channels around the velocity shown in the panel. In this region we find the strongest GMCs in M\,33, the triangle is the strongest cloud found by \citet{Engargiola03} and the star is NGC604, a giant HII region. The strongest clouds have a velocity width of $\sim$10\kms.}
   \label{channel}
   \end{figure*}

The zero-order moment of the cube was calculated to obtain the integrated CO intensity map (Fig.~\ref{I_CO}). In such a large map the line position varies according to the rotation curve. It is necessary to take into account this variation in the intensity map. Given that we wanted to be sensitive to CO emission not centered at the same velocity as the HI, we used the HI cube to determine a rough rotation curve and then integrated the CO over that range plus a few \kms\ to either side.  Any warp or other feature that affects the gas will necessarily be included. Thus we defined three velocity windows : -210 to -260\,\kms, -250 to -280\,\kms, and -255 to -282\,\kms. Complexes around $\alpha$=\hms{01}{34}{33.2}, $\delta$=\dms{30}{47}{5.6} (corresponding to the HII region NGC\,604) and $\alpha$=\hms{01}{34}{04}, $\delta$=\dms{30}{54}{56} have their own velocity range from -205 to -260\,\kms, and -247 to -270\,\kms. The noise level strongly depends on the position in the map; the area with $\delta>$30\degre 58\arcmin\ was scanned much longer than the South where the emission is strong. We used a detection threshold of 3$\sigma$ where $\sigma$ is the rms noise outside of the velocity windows as defined above. The three parts in Fig.~\ref{I_CO} correspond to the velocity windows  defined above. Each part has a different color scale because the line intensity varies from 13\,K \kms, for the strongest cloud in the bottom part, to less than 0.5\,K\,\kms\ in the upper part. We use a second intensity map where all channels (in velocity windows) are summed without a threshold -- this map is used to derive the flux, mass surface density and azimuthal average.  A strictly positive map, such as based on a 3$\sigma$ detection threshold, necessarily includes noise which happens to be above the threshold without subtracting the negative noise peaks. We believe this is responsible for the nearly constant positive CO emission in the \citet{Heyer04} radial profile at large radii.

\section{Molecular emission}

In this section we present the distribution of the CO emission over the area we observed. The emission is not uniformly distributed over the area. We distinguish three parts in the map shown in Fig. ~\ref{I_CO}. The strongest clouds, observed by \citet{Engargiola03}, are found in the southern part. In the middle part, the CO intensity decreases by a factor 2-3 and we detect the furthest cloud of \citet{Engargiola03} (their tentative detection at $\alpha$=\hms{1}{34}{16}, $\delta$=\dms{30}{52}{21}) and many clouds not detected by them.  We detect CO emission in the North out to $\alpha$=\hms{01}{34}{16.1}, $\delta$=\dms{31}{03}{58.0}.

In the southern part of the map, CO emission traces an inner arm where we detect the strongest clouds of M\,33. The strongest cloud is represented by a triangle on Fig.~\ref{channel}. We detect the clouds found by \citet{Engargiola03} (see their catalog) as well as many other clouds and more diffuse emission. While we observed the CO(2--1) transition, the sensitivity of HERA permits us to see diffuse emission in the arm, probably missed by \citet{Engargiola03} because the interferometer resolves out the large scale emission.

Above the main arm we see two large complexes at $\alpha$=\hms{01}{34}{16}, $\delta$=\dms{30}{52}{22} and $\alpha$=\hms{01}{33}{59},$\delta$=\dms{30}{55}{00}. These complexes have radii of 35\arcsec\ and 75\arcsec and each have several emission peaks surrounded by diffuse emission. The distance between peaks is about 30\arcsec\ and velocity peaks differ from 2 to 5\,\kms\ (i.e. 1 or 2 channels). The large complex at \hms{01}{33}{59}; \dms{30}{55}{00}, which has a circular shape, will be the subject of a future more detailed treatment. Up to 30\degre58\arcmin\ molecular clouds are clearly identified because they are resolved by the beam. In the arm structure at $\delta$=\dms{30}{56}{30} the size of clouds is similar to southern clouds (R$_{\mathrm{cl}}\sim$15 - 30\arcsec) but the intensity decreases to 0.7\,K\,\kms. One of the interesting results of these observations is just how much the cloud properties change over the map.

\begin{figure}[!ht]  
   \centering
   \includegraphics[width=250pt]{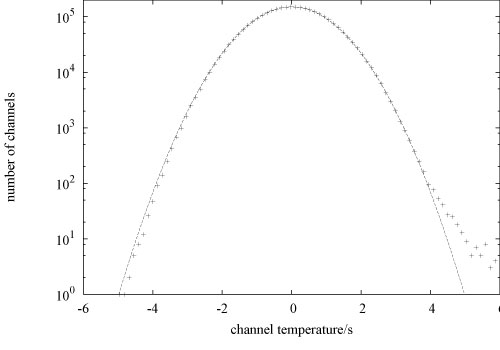}
   \includegraphics[width=250pt]{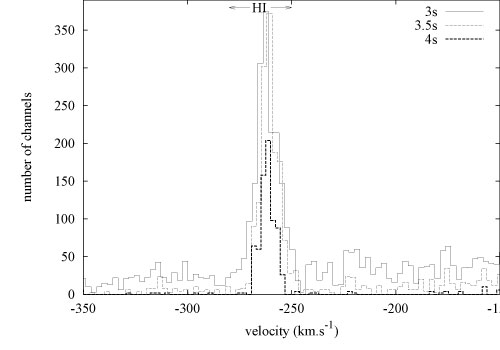}
   \caption{\textit{Upper panel} : Number of channels in the upper part of the cube ($\delta>$30\degre58\arcmin) as a function of the ratio between the temperature of a channel and the noise level for that spatial pixel. Only channels in the velocity range -400\,\kms\ to -100\,\kms\ were included. The deviation with respect to the gaussian noise distribution (full line) is due to the CO emission. \textit{Lower panel} : Number of channels above 3, 3.5 and 4 $\sigma$ (where $\sigma$ is the rms noise level) as a function of the velocity in the upper part of the cube. The 3.5$\sigma$ and 4$\sigma$ curve are multiplied by 4. Many more peaks are present within the velocities expected from the HI rotation curve, showing that signal is present at these velocities.}
   \label{noise_distribution}   
   \end{figure}

\begin{figure}[ht]
   \centering
   \includegraphics[width=250pt]{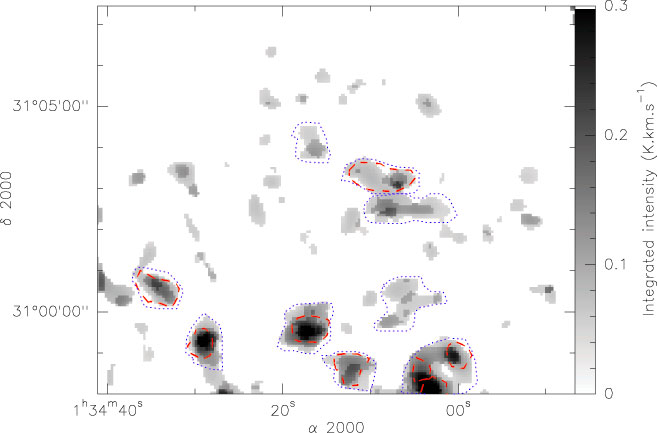}
   \caption{Northern half of the integrated intensity map for the 30\arcsec\ smoothed cube. As the noise level decreases by a factor $\sim$2 with the smoothing we are able to identify fainter emission. Using the 3$\sigma$ and the two channel velocity criterion we identify the cloud at $\alpha$=\hms{01}{34}{16.1}, $\delta$=\dms{31}{03}{58.0} to be our furthest detection. Dashed lines correspond to clouds identified in the 15\arcsec\ resolution map and dotted contours are clouds or groups of clouds identified in the 30\arcsec\ map. Nearly half of the mass in this part of the galaxy is contained in these clouds.}
   \label{smooth_intensity}
   \end{figure}

In the northernmost part, CO emission is weak and not easily disentangled from the noise. We scanned the northern half of the cube ($\delta>$30\degre58\arcmin) channel per channel in order to look for emission. Figure~\ref{noise_distribution} (upper panel) represents the number of channels as a function of the signal to noise ratio. The noise distribution follows a gaussian distribution with a deviation at levels above $\sim$3$\sigma$ showing the existence of molecular emission. Figure \ref{noise_distribution} (lower panel) shows the number of channels (in the northern part) above 3, 3.5 and 4$\sigma$ as a function of the velocity. Many more peaks are present between -275\,\kms\ and -250\,\kms, as expected from the HI rotation curve. However a single channel peak is not necessarily a detection. Our criteria for detection is that the signal must be greater than 3$\sigma$ over (at least) two adjacent velocity channels. The velocity peak should fall within the HI velocity range. Using these criteria, the furthest detection in our map is found at $\alpha$=\hms{01}{34}{16.1}, $\delta$=\dms{31}{03}{58.0}. The integrated intensity of the 30\arcsec\ smoothed cube (Fig. \ref{smooth_intensity}) shows the spatial distibution of this faint emission. 

We identify 9 clouds on the 30\arcsec\ resolution map (dotted contours on Fig. \ref{smooth_intensity}). For comparison we see 7 clouds on the 15\arcsec\ map (dashed contours). We see a large but faint complex at $\alpha$=\hms{01}{34}{08}, $\delta$=\dms{31}{02}{55}. The comparison with HI velocities and the spatial distribution of the atomic gas and star formation confirms the detection of CO in the North. The molecular mass (H$_{2}$ + associated He) is determined from the CO intensity as 
\begin{eqnarray}
 \label{MCO}
   \mathrm{M}_{H_{2}}=\sum_{pixels} 2m_{p}\times1.36 \times X\times\int\mathrm{T(v)dv} \times\delta x \delta y D^{2}
\end{eqnarray}
where T(v) is the temperature of the channel at the velocity v and $\delta x \delta y D^{2}$ is the projected surface of the pixel in the sky. m$_{\mathrm{p}}$ is the mass of the proton and the factor 1.36 corresponds to the helium fraction. We used a ''standard'' \citep[e.g.][]{Dickman86} CO-to-H$_{2}$ conversion factor X=2\dix{20}\,\Xunit. This conversion factor is given for the CO(1--0) line and we assume a line ratio R$_{12}$=CO(2--1)/CO(1--0) to use it for the CO(2--1) line.
In spiral galaxies the line ratio varies from 0.65 for M\,31 \citep{Nieten06} to 0.80 for M\,51 \citep{Garcia-Burillo93}. \citet{Thornley94} determine a mean ratio of 0.70 over M\,33. For the Galaxy, R$_{12}$=0.7-0.8 \citep{Sakamoto97}. We adopt a mean value of 0.75 such that $\ratiob =$ 2.7\dix{20}\,\Xunit. \citet{Lundgren04} find a line ratio of 0.77 for M\,83 with an arm/inter-arm variation of $\sim$25\%.

Summing the CO intensity over the whole moment-zero map (with no detection threshold) and considering for the moment the CO-to-H$_{2}$ conversion factor to be constant, we estimate the H$_2$ mass to be 1.6\dix{7}\msol\ within the region we mapped. Most of the mass ($\sim$90\%) is contained in  the southern half of the map. In the North ($\delta>$30\degre58\arcmin) we detect a total mass of 9$\pm$2\dix{5}\,\msol. Summing the intensity within the nine polygons defined for the identified clouds we obtain 3.5$\pm$1\dix{5}\,\msol, i.e. about the third of the mass of this region. The mean noise level in the 30\arcsec\ map is 7\,mK. Integrating the one sigma noise level over the pixels of the northern part yields a mass of 4.5\dix{5}\,\msol. Is this the trace of diffuse emission? Clouds smaller than the beam could produce such emission.

\section{Radial distribution}
\label{radial}

\begin{figure}[!t]
   \centering
   \includegraphics[width=250pt]{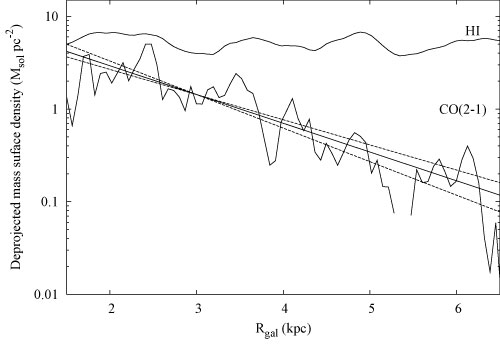}
   \caption{Azimuthal average of the deprojected H$_{2}$ (+He) mass surface density. The CO emission is averaged over ellipses 16\arcsec\ in width and expressed in molecular mass surface density using equation~\ref{MCO}. The HI average is computed over our CO area. The radial profile is fitted by an exponential disk from 3.5 to 6.5\,kpc, with a scale length of 1.4$\pm$0.1 kpc. Dashed lines show the estimated uncertainties on disk parameters. The HI (+He) profile is roughly constant with a mass surface density of 5\,\mpc. HI emission peaks coincide with CO peaks, which tends to show that CO -and \H2- forms on HI. The CO peak at 4\,kpc is not correlated with HI, this cloud is an inter-arm cloud which is not associated with star formation tracers.}
   \label{I_CO_r}
   \end{figure}

We derive the azimuthal average of the CO emission from the (total) integrated intensity map. We average the emission on rings inclined by i=56\degre\ and tilted of PA=23\degre. Each ring has a deprojected width of 16\arcsec, $\pm$8\arcsec\ around the nominal galactocentric radius of the ring. The CO surface brighness is deprojected to face-on. Our map covers $\sim$20-25\% of the total ellipse where R$_{gal}$=4\,kpc, and 10-20\% for greater radii. Here we probe about 20\% of the disk with a variation of radii from 1.3 to 8\,kpc ($\sim$R$_{25}$). In Fig.~\ref{I_CO_r} we show the azimuthal average in deprojected H$_{2}$ mass surface density calculated using equation (\ref{MCO}). 

In the radial profile we see the three parts mentioned in the previous section. The southern part and the middle part correspond to plateaux at 1.5\,kpc and 2.5\,kpc. In the main arm (South) the mean value of the (deprojected, i.e. face-on) \H2 mass surface density is about 3\,\mpc\ and drops to 1\,\mpc\ in the middle part. At larger radii, from 3.5 to 7\,kpc, the decrease in intensity is more regular. Beyond 7 kpc, too few pixels are available to calculate the azimuthal average. From 3.5 to 6.8\,kpc the \H2 mass surface density decreases to 0.1\,\mpc. We fit the azimuthal average with an exponential disk  between 3.5 and 7\,kpc. We do not take into account the range from 1.5 to 3.5\,kpc because few pixels represent these radii and most of the emission is due to the big clouds in the main arm. The best fit gives a radial scalelength of R$_{\mathrm{CO}}$=1.4$\pm$0.1\,kpc (Fig.~\ref{I_CO_r}). Because our data poorly sample the inner part and indeed we are interested in the outer part, we fit exponential disks by eye to the data from 3.5 kpc onward, avoiding the sharp rise due to the most intense clouds in the main arm. As can be seen, these fits ended up fitting the inner regions as well.  To estimate the uncertainty in the scale lengths, other slopes were plotted (as in Fig.~\ref{I_CO_r}) designed to be the extreme values the scale length parameter could take. R$_{\mathrm{CO}}$=1.4\,kpc is compatible with the radial decrease found by \citet{Engargiola03} between 0 and 3\,kpc. Figure 1 of \citet{Heyer04} shows that the CO emission from the main arm (R$\sim$2.5 kpc) creates an asymmetry favoring the North but that at larger radii (R$\sim$4 kpc) the southern emission is stronger. However, comparison of our map (Fig. 2) with their Fig. 1 shows that they missed emission present in the North at radii around 4\,kpc. An exponential disk wih a radial scalelength of 1.4\,kpc fits the CO integrated intensity over both the inner (R$_{gal}<$3\,kpc) and the outer parts of M\,33.

Integration of the exponential disk gives an estimate of the total \H2 mass of M\,33: $M_{H_{2}}=\Sigma_{0}2\pi R_{CO}^{2}$. The best fit to the mass surface density gives the deprojected central density $\Sigma_{0}$ of 10$\pm$1\,\mpc. Thus we find a total mass of \H2 : M$_{H_{2}}$=1.2$\pm$0.3\dix{8}\,\msol. This value is higher by a factor 2-3 than \citet{Engargiola03} although they corrected their mass estimate to take into account the mass lost through interferometric filtering. We detect more CO emission in the same area and thus calculate a higher H$_2$ mass. Since the cloud brightness decreases with galactocentric distance, and they only observed the central region with a single dish for comparison, it could be that a growing fraction of the clouds fall below the sensitivity of their interferometer observations. Thus, since their mass estimate is based on multiplying the catalogued clouds by a factor two, they necessarily exclude flux from areas where they detect no clouds. With their new observations \citet{Rosolowsky07} give a better correction of the mass and find $M_{H_{2}}$=1.2\dix{8}\,\msol. \citet{Heyer04} find a higher mass than we do but we believe they overestimate the radial scalelength of the CO emission. They used a masked moment map to derive the CO intensity from their cube, hence they take into account only the positive emission and find a longer scalelength (and thus larger total H$_2$ mass, calculated using the parameters of the CO exponential disk).

\section{Comparison with the HI}
\label{HI}
Figure~\ref{contours} (bottom left) shows the CO integrated intensity map of the 30\arcsec\ resolution cube with HI contours from a cube at the same angular resolution \citep{Deul87}. In the southern part the CO emission fills the lower HI contour in the main arm. Up to 30\degre 57\arcmin\ we see CO emission associated with HI column density peaks.  With a few exceptions, the molecular clouds lie on or close to HI peaks or ridges above a column density of $\sim$1\dix{21}\,cm$^{-2}$. HI peaks are not necessarily associated with CO emission but the CO emission is almost always associated with HI peaks, the exceptions being clouds at ($\alpha$=\hms{01}{34}{25}, $\delta$=\dms{30}{54}{50}) and ($\alpha$=\hms{01}{34}{16.9}, $\delta$=\dms{30}{59}{31.4}). They are discussed in the next section.

\begin{figure}[!ht]
    \centering
    \includegraphics[width=250pt]{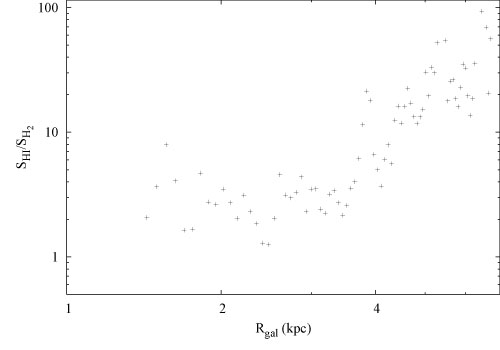}
    \caption{Ratio of the HI to the H${_2}$ deprojected surface densities as a function of galactocentric radius (i.e. azimuthally averaged). In the main arm (R$_{\mathrm{gal}}<$3.5\,kpc) the ratio is roughly constant at $\sim$3 and then increases to $\sim$30 in the outer parts.}
    \label{HI_H2}
    \end{figure}

In Fig.~\ref{I_CO_r} we show the radial distribution of the HI in the area observed in CO, calculated as in section~\ref{radial}. The azimuthally averaged mass surface density of the HI is roughly constant ($\sim$5\,\mpc) over our zone. However, the small bumps in the radial distribution seem correlated with sharper bumps in the CO distribution. Figure~\ref{HI_H2} gives the ratio of the azimuthal average of the mass surface densities $\Sigma_{HI}$ to $\Sigma_{H_2}$. Below 3.5 kpc the ratio is roughly constant, with a value of $\sim$2-3, although the \H2 surface density decreases by a factor $\sim$5 between 1.5 and 3.5\,kpc. A break in the ratio occurs at 3.5 kpc, where the ratio starts to increase sharply and reaches $\sim$20-30 at 5.5\,kpc. Beyond 5.5\,kpc the uncertainties in the CO emission lead to a high dispersion in the $\Sigma_{HI}$ to $\Sigma_{H_2}$ ratio. Inspection of Heyer et al's Fig. 4 shows rough agreement, although because they had many fewer points in the 1.5--3.5\,kpc range, they traced a single power law $\Sigma_{HI} / \Sigma_{H_2} = 0.9 R^{0.6}$. All our analyses of H$_2$ mass surface densities use the CO integrated intensity map made {\it without} a detection threshold -- in this way positive and negative noise cancel out.

The spiral structure of M\,33 is not well defined and any density wave is weak. The stellar arms seen on the Blue band image (Fig.~\ref{CO_box} left panel) are not as contrasted as the HI arms and the spiral structure of \citet{Regan94} is essentially within the first 1.5\,kpc. The correspondance with old stellar emission is poor within our zone. The HI intensity map shows a filamentary structure not seen in the optical/NIR bands. In M\,31 \citep{Nieten06}, for example, the arm-interarm contrast in CO emission, and thus presumably H$_2$ surface density, is much higher than that of the stars or atomic gas. The \citet{Engargiola03} results support the role of spiral arms in molecular cloud formation and indeed the CO emission in our map (Fig.~\ref{contours})  is very well correlated with the spiral arms in M\,33 as traced by star formation as seen in the UV, \Ha, HI, or thermal IR emission. It is often suggested that the transformation of atomic into molecular gas occurs when the atomic gas is compressed upon entering a spiral arm. This is probably not the only formation mechanism as \citet{Braine04}  found CO emission well beyond any identifiable spiral arms in the Sc spiral NGC\,4414. In M\,33, the gas surface density is dominant (over the stellar component) beyond about 3.5\,kpc from the center so a stellar density wave is certainly not necessary for the formation of molecular gas. Furthermore, we find one major cloud in a very clear interarm region and not associated with tracers of star formation, hence the name "Lonely Cloud".

\section{The lonely cloud, a CO bright interarm GMC}
 
The coordinates of the CO maximum of the interarm cloud are $\alpha$=\hms{01}{34}{16.9}, $\delta$=\dms{30}{59}{31} and the cloud is easily seen in Fig.~\ref{I_CO}. The HI line is detected at this position but is wide and lacking a clear peak and the HI column density is low compared to the adjacent spiral arms. The Lonely Cloud is one of the CO brightest clouds far out in the disk of M\,33 with a peak main beam temperature of about 250\,mK in the CO(2--1) line. What is surprising about this cloud is that there is no detected \Ha\ or FUV emission and the 8\mum\ et 24\mum\ show only very weak emission, yet the CO is quite strong. We would never have detected this cloud without a blind search and one of the advantages of mapping a substantial region with HERA is that these "odd" clouds can be detected.

Over an area about (70\,pc)$^2$, roughly that of a standard GMC, the molecular gas mass is about  2\dix{4}\,\msol, using a "standard" $\ratioo$ factor as before.  This is equivalent to the HI mass over the same area. However, the $\ratioo$ factor is probably somewhat higher in this region for two reasons: ($i$) the interstellar radiation field is very weak, such that the cloud is likely cool and ($ii$) the $\ratioo$ factor is generally believed to increase with galactocentric distance, possibly due to a metallicity dependance. It is thus likely to be predominantly molecular despite being in a clear interarm region.  The stellar surface mass density is quite low (below that of the HI) at this position, some 5 kpc from the center of M\,33.  No obvious heating sources other than the ambient radiation field or mechanisms internal to the cloud (such as gravity) are present and yet the antenna temperature is about 250 mK in the CO(2-1) line.  This shows that even far out in spiral disks where the stellar mass contribution is less than that of the gas, the molecular gas is not so cold that it is not detectable or even difficult to detect as long as the spatial resolution is sufficient to resolve the cloud. These heating sources and cosmic ray heating are sufficient to make the CO visible despite the low metallicity. The CO emission can thus be used to trace the H$_2$, although likely with a $\ratioo$ factor somewhat higher than in the molecular ring of the Galaxy.  

At about $\alpha$=\hms{01}{34}{25.7}; $\delta$=\dms{30}{54}{48}, another cloud not associated with strong HI emission is present. The emission at 8\mum\ and 24\mum\ is not very strong but stronger than for the "Lonely" cloud and FUV and \Ha\ emission are detected, although principally associated with the western side of the cloud. The molecular fraction is greater than unity in this low HI column density zone but some star formation is taking place. The molecular gas mass and surface density are similar to those of the Lonely cloud and the HI column density is equally low.
 
\section{CO as a tracer of star formation}

\begin{figure*}[!ht]
 \centering
 \includegraphics[width=450pt]{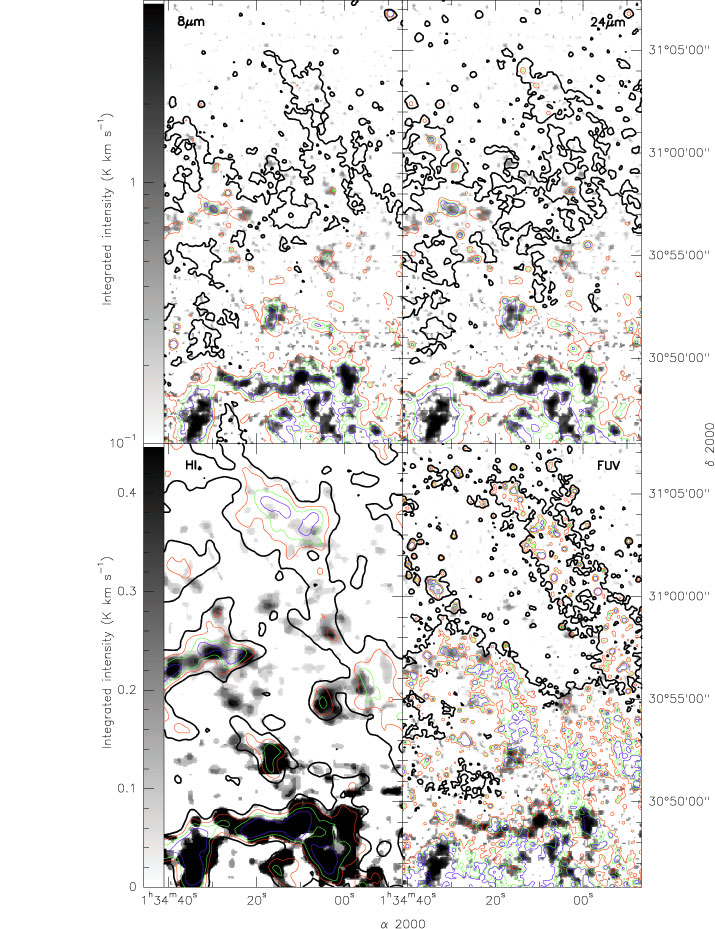}
 \caption{HI (Deul \& Van der Hulst, 1987), FUV (GALEX at 160\,nm), 8\mum\ and 24\mum\ (SPITZER) contours on the CO intensity map. The FUV and IR maps are smoothed at the resolution of the CO map (15\arcsec). The CO map is convolved to 30\arcsec\ resolution in the lower left panel to compare with the HI. The CO grey scale is identical for the 8\mum, 24\mum\ and NUV comparison. Contours are for HI from 9\dix{20}\,cm$^{-2}$ with steps of 3\dix{20}\,cm$^{-2}$, for FUV from 4.3\,$\mu$Jy\,arsec$^{-2}$ with steps of 6.5\,$\mu$Jy\,arsec$^{-2}$, 8\mum\ contours are 0.5\,MJy\,sr$^{-1}$ with steps of 0.5\,MJy\,sr$^{-1}$ and 24\mum\ are from 0.3\,MJy\,sr$^{-1}$ with steps of 0.6\,MJy\,sr$^{-1}$. The bold line represents the first contour above the noise. In each band, emission peaks correspond to strong CO emission up to $\delta$=30\degre 57\arcmin. The CO emission, as the FUV and IR emission, is confined to the HI arms except for a cloud at $\alpha$=\hms{01}{34}{16.9}, $\delta$=\dms{30}{59}{31.4}. This inter-arm cloud is not associated with FUV or IR peak, and is in a low HI column density region.}
 \label{contours}
\end{figure*}

In this section we compare the CO emission with the  emission at 8\mum, 24\mum, 160\,nm, and \Ha, all of which trace star formation. Figure~\ref{contours} shows the CO emission in greyscale with contours of 8\mum\ PAH emission (top left), 24\mum\ dust emission (top right), and 160\,nm UV emission (bottom right). The UV image is from the GALEX FUV band observations at 160\,nm \citep{Thilker05}. The initial resolution of $\sim$5\arcsec\ is degraded to the resolution of the CO observations. The 8\mum\ and 24\mum\ images are from Spitzer (IRAC band 4 and MIPS band 1). The background subtraction in these images is crucial for our purposes because of the extremely low level of the emission in the outer parts of the galaxy. In the 8\mum\ map the background is defined by averaging small areas at the edge of the map, avoiding obvious stars. The background value is estimated to be 4.49\,MJy\,sr$^{-1}$ in the image as retrieved from the archive. At 24\mum\ the background shows a substantial gradient in the scan direction. To eliminate the gradient, we selected groups of pixels with no apparent signal and far from any regions where signal is expected based on data at other wavelengths -- these are taken to represent the background. A plane is then fit using a least square minimization through these pixels and subtracted from the whole map. Two independent MIPS 24\mum\ maps were treated separately and then combined.

\begin{table}[!b]
\begin{center}
\caption{List of the tracers used to compare with the CO emission. We determine the scalelength for the area we observed and errorbars are estimated as explained for the CO.}
\label{tracers}
\begin{tabular}{lcc}
\hline
tracer       & R$_{d}$ (kpc)    & reference \\
\hline
\hline
CO (1-0)     & 2.0              & \citet{Heyer04} \\
CO (1-0)     & 1.4$\pm$0.1      & \citet{Engargiola03} \\
CO (2-1)     & 1.4$\pm$0.1      & this paper \\
\hline
UV (160\,nm)  & 2.10$\pm$0.10   & GALEX \citep{Thilker05} \\
H$_{\alpha}$ & 1.80$\pm$0.15    & \citet{Hoopes00} \\
K band       & 1.00$\pm$0.05    & 2MASS $^{*}$\\
8\mum        & 1.44$\pm$0.05    & Spitzer \\
24\mum       & 1.55$\pm$0.10    & Spitzer \\
60\mum       & 1.30$\pm$0.05    & ISO \citep{Hippelein03} $^{*}$  \\
170\mum      & 1.80$\pm$0.10    & ISO \citep{Hippelein03}  \\
HI           & $\sim$constant   & \citet{Deul87} \\
\hline
\end{tabular}
     
 \begin{list}{}{}
  \item[(*)] \citet{Regan94} obtain longer scale lengths (R$_{K}$=1.4\,kpc and R$_{60\mu m}$=1.5\,kpc).  The difference is due to the fact that they determine the scale length over the whole galaxy whereas the scale length given above is for the area we observed in CO.
 \end{list}
\end{center}
\end{table}

In the southern part of the map UV, IR and CO emission fall within the HI arms (main arm and filament). Up to 30\degre 57\arcmin\ GMCs coincide with regions of active star formation. The 8\mum\ and the 24\mum\ emission, which trace respectively PAH emission and hot dust and/or very small grains, have the same scalelength as the CO (within the uncertainties see Tab.~\ref{tracers}). The dust emission per H-atom is proportional to the metallicity because the grains are mostly composed of metals so the dust mass varies with Z (metallicity as a mass fraction) while the metallicity does not affect the gas mass \citep{Mezger90}.

\begin{figure*}[!ht]
    \centering
    \includegraphics[width=350pt,angle=270.]{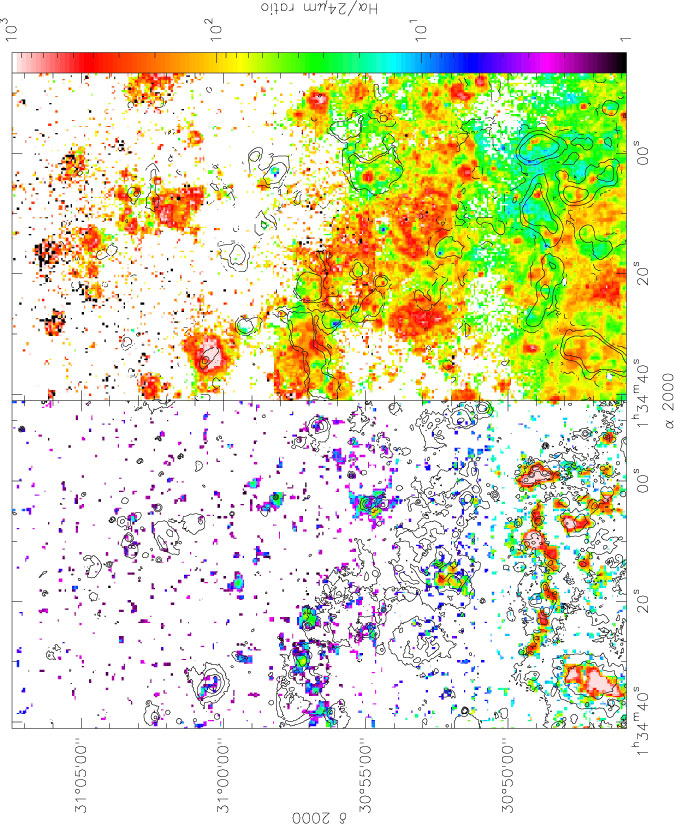}
    \caption{CO map with the \Ha\ \citep{Hoopes00,Greenawalt98} contours (left) and map of the \Ha\ to the 24\mum\ ratio (right). \Ha\ contours are 40, 160 and 640 \dix{-18} erg\,cm$^{-2}$\,s$^{-1}$\,arcsec$^{-2}$. Contours on the ratio map are contours of the 30\arcsec\, resolution CO map, and levels are 2, 1, 0.2, 0.1 and 0.05 K\,\kms. The ratio is calculated where the \Ha\ brightness is above 12\dix{-18} erg\,cm$^{-2}$\,s$^{-1}$\,arcsec$^{-2}$ but the result does not depend on this cutoff. The \Ha\ to 24\mum\ is roughly constant up to 30\degre 58\arcmin\ both in clouds and inter-arm medium. Above the second arm the 24\mum\ emission decreases faster the \Ha\ (R$_{24\mu m}<$R$_{H_{\alpha}}$). The black pixels in the map of the \Ha / 24\mum\ ratio are due to negative 24\mum\ values.}
    \label{halpha}
    \end{figure*}

\begin{figure}[!ht]
    \centering
    \includegraphics[width=250pt]{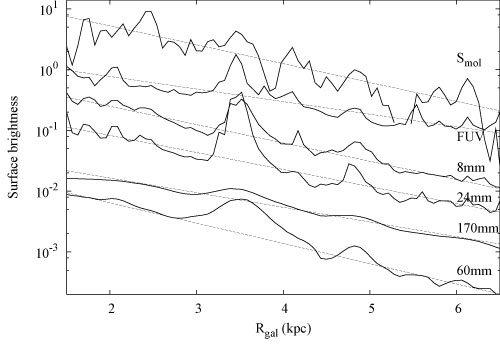}
    \caption{Radial profile of various star formation tracers in the region of our CO map. Fluxes are normalized to the deprojected CO mass surface density. The solid lines show exponential disk fits for each tracer. The fit does not take into account high emission peaks (like the peak at 3\,kpc). Most of the structures in the CO profile are correlated with star formation tracers while the radial decrease changes. The scale length of the CO emission is lower than the UV and \Ha. The CO and 8\mum\ emission have the same scale length R$_{8\mu m}$=1.4\,kpc. The 24\mum\ emission scale length is R$_{24\mu m}$=1.6\,kpc.}
    \label{average}
    \end{figure}

At large scales, there is an excellent spatial correlation between the CO emission and star formation tracers, with the exception of the Lonely cloud. However, the ratios between CO and the other tracers of star formation vary from cloud to cloud and in a general way with galactocentric radius. Figure~\ref{halpha} (left) shows the CO map with \Ha\ contours. In the main arm the \Ha\ emission is mainly associated with the CO emission but there are also zones with \Ha\ but little or no CO emission. Figure~\ref{average} presents the azimuthal average of the star formation tracers averaged the area of the CO map, such that the azimuthal averages are all based on the same physical regions. The averages are expressed in surface brightness and normalized to the CO. Solid lines show exponential fits to the azimuthal averages. The fits do not take into account strong emission peaks such as the peak at 3.5\,kpc due to NGC\,604. Table~\ref{tracers} gives the radial scale length for each of the tracers. Error bars are estimated as explained in section~\ref{radial} for the CO.

Figure~\ref{halpha} (right) shows the \Ha\ to 24\mum\ ratio in arbitrary units over the area observed in CO. Only pixels with an \Ha\ intensity greater than or equal to 1.2\dix{-17}\,erg\,s$^{-1}$\,arcsec$^{-2}$ are included -- other areas are white. In the main arm, the CO-bright regions have a rather constant \Ha/24\mum\ ratio of $\sim 35$ (green). Generally, in the CO-rich southern zones, the \Ha/24\mum\ ratio is lower than where little molecular gas is present. In some molecular clouds further North the ratio is still similar but there is a clear change to higher \Ha/24\mum\ intensity ratios as the outer parts are reached. At large scales, the change in the \Ha/24\mum\ ratio is close to an order of magnitude, going from a dominant green-yellow in the South to red in the North. Despite this change, their spatial correlation is excellent and the figure is unchanged if the noise threshold is applied to the 24\mum\ map. The near-identical appearances of the \Ha\ and 24\mum\ images, which both have a resolution of 5 -- 10\,pc, is a powerful illustration that the star formation in M\,33 is visible rather than being dominated by deeply embedded sources. This had been noted for M\,31 by \citet{Devereux94} at arcminute scales and by \citet{Rodriguez-Fernandez06} at arcsecond scales within the outer arm. The increase in the \Ha/24\mum\ ratio may be explained by the change of the metallicity. If the metallicity decreases the \Ha\ is less affected by the extinction. At the same time the amount of dust that emits at 24\,\mum\ decreases. Both effects increase the \Ha\ to 24\,\mum\ ratio.

\section{The Star Formation Efficiency}

What is the star formation rate and how efficient is the transformation of gas into stars in M\,33? From the \Ha\ observations made by \citet{Hoopes00} and \citet{Greenawalt98}, we can estimate the star formation rate. The stellar continuum has been subtracted and the remaining stars masked. \citet{Kennicutt98} gives the relation between the \Ha\ luminosity and the star formation rate 
\begin{eqnarray}
 \begin{centering}
  SFR (M_{\odot}\,yr^{-1}) = \frac{L_{H_{\alpha}}}{1.26\times 10^{41}\mathrm{(erg \, s^{-1})}} = 3 \times 10^{-8} \frac{L_{H_{\alpha}}}{(L_\odot)}
 \end{centering} 
\end{eqnarray}
where $L_{H_{\alpha}}$ has been corrected for extinction, \citet{Kennicutt98} assumes 1.1 mag (factor 2.8) of extinction. \citet{Devereux97} give a mean extinction for HII regions of A$_{\mathrm{V}}\sim$1\,mag, without variation with the radius. However the global correction for the \Ha\ flux is about A$_{\mathrm{V}}\sim$0-0.4\,mag. Hence we consider the extinction to be between 0 and 1\,mag. The flux we measure for the disk of M\,33 is 3\dix{40}\,erg s$^{-1}$ or 7.9\dix{6}\,L$_\odot$ and 20\% of this flux is found within the area we observed in CO. We obtain a star formation rate of $0.05 \la SFR \la 0.13$ over our map and 5 times more for the entire galaxy. This range coincides well with estimates from a variety of observations \citep{Blitz06,Kennicutt98,Hippelein03}, which vary from 0.26 to 0.7\,\msol \,yr$^{-1}$ for the entire disk.

We can also estimate the star formation rate using the NUV observations from GALEX \citep{Thilker05}. The conversion between L$_{UV}$ and the star formation rate from \citet{Kennicutt98b} is
\begin{eqnarray}
 \begin{centering}
  SFR (M_{\odot}\,yr^{-1}) = L_{UV}\times 1.4\times 10^{-28}\mathrm{(erg \, s^{-1}\,Hz^{-1})}
 \end{centering} 
\end{eqnarray}
where L$_{UV}$ is the luminosity over the wavelength range 1500-2800\,$\AA$. The GALEX NUV band is 1770-2830\,$\AA$ so that the Kennicutt relation could be applied. Within our area we mesure a luminosity of L$_{NUV}$=7.0\dix{27}\,erg\,s$^{-1}$\,Hz$^{-1}$ which leads to a SFR of 0.1\,\msol \,yr$^{-1}$, not corrected for extinction. The SFR derived from the NUV luminosity is in good agreement with the SFR from the \Ha\ luminosity. The \Ha\ emission is from very young stars surrounded by HII regions whereas the UV emission comes from (on average) a slightly older stellar population.  As such, we expect the \Ha\ to be in regions of higher extinction (closer to the molecular clouds) than the UV.  \citet{Devereux97} find the extinction towards HII regions to be 1\,mag and several times less elsewhere.

From a total \H2 (+He) mass within our map of 1.6\dix{7}\,\msol, the molecular depletion timescale is then 1.1 -- 3.2\dix{8}\,yr. {\it M\,33 is then significantly more efficient at forming stars than large local universe spirals, much like what is observed at intermediate redshifts.} The \citet{Kennicutt98} sample of "normal" spirals yields a molecular gas depletion timescale, using our $\ratioo$ factor and including He, from 0.9 to 2.5\,Gyr depending on the extinction correction.  Excluding the small galaxies (M\,33, NGC\,2403, NGC\,1569) increases the time. This is an order of magnitude greater than what we find for M33 (0.11 -- 0.32\,Gyr). In their study comparing CO and radio continuum emission, which they use to trace the star formation rate, \citet{Murgia02} find a molecular depletion time of 2.7\,Gyr. Why would M\,33 be more efficient in star forming than normal spiral galaxies? How certain can we be of this high SFE?\\
($i$) The SFR within our map is certainly at least 0.046\,\msol \,yr\,$^{-1}$, from the \Ha\ flux and assuming no extinction. We can also use the 60\mum\ map from \citet{Hippelein03} to estimate the fraction of the star formation within our map -- 14\% of the total flux is within our map but we expect the 60\mum\ flux to underestimate the star formation in the outer parts because of the metallicity and temperature gradient which yield  a very short 60\mum\ scale length.  At 170\mum\ the fraction would be higher (Tab. 2) but the map does not cover enough of the galaxy to make a direct measurement. The average of the whole-disk values of the SFR from the litterature yields $0.5\pm 0.2$\,\msol \,yr$^{-1}$, such that dividing by 5 (from the \Ha\ and probably an underestimate due to the extinction variation) or 7 (from the 60\mum\ and likely an overestimate due to the metallicity gradient) yields an SFR in our map of 0.05 -- 0.1\,\msol \,yr$^{-1}$, at least the 0.046\,\msol \,yr$^{-1}$ obtained from the \Ha\ assuming no extinction.\\
($ii$) The other element is the $\ratioo$ ratio, which could be higher in M\,33, which has a slightly subsolar metallicity. 90\% of the CO flux is in the lower part of our map, at galactocentric radii sampled by other observers or only a bit further out and where any increase in the $\ratioo$ factor due to the metallicity should be weak. \citet{Engargiola03} use a conversion factor of $\ratioo$=2\dix{20}\,\Xunit\ as we do. \citet{Blitz07} suggests $\ratioo$=3\dix{20}\,\Xunit, possibly ranging up to $\ratioo$=4.5\dix{20}\,\Xunit. \citet{Wilson95} suggests $\ratioo$=1.5\dix{20}\,\Xunit\ for the central regions and about $\ratioo$=4\dix{20}\,\Xunit\ further out. However, a much more radical increase in the $\ratioo$ factor is necessary to bring the molecular depletion time scale in M\,33 up to the average value for spirals.

\begin{figure}[!t]
     \centering
     \includegraphics[width=250pt]{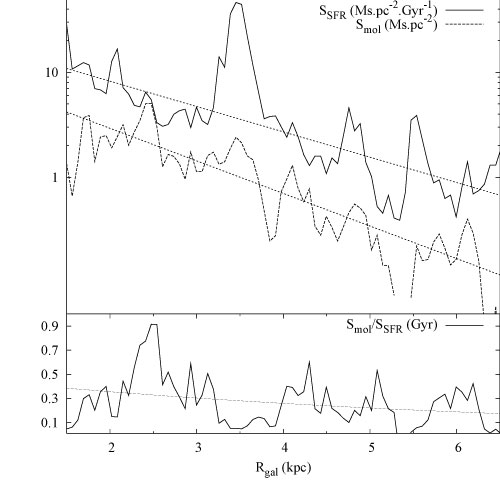}
     \caption{Azimuthal average of the \Ha\ emission converted into a star formation rate using the Kennicutt relation and the deprojected H$_{2}$ mass surface density. Dotted lines are the best fit of an exponential disk with a scale length of 1.8\,kpc for \Ha\ and 1.4\,kpc for the CO. The lower panel gives the ratio of the surface density of molecular gas $\Sigma_{\mathrm H2}$ to the Star Formation Rate surface density. The dashed line corresponds to the ratio of the exponential fits, showing that the depletion timescale decreases with radius. However, the cloud to cloud variation is higher than the decrease itself.  Is this due to seeing the clouds in different evolutionary phases ?}
     \label{SFR}
     \end{figure}

Figure~\ref{SFR} (upper panel) gives the azimuthal average, over the area we mapped in CO, of the \Ha\ intensity converted into SFR (using the Kennicutt 1998b relation). The best fit to the SFR decrease has a scalelength of 1.8$\pm$0.1\,kpc. As for the other tracers of star formation, the fit does not take into account the strong peak at 3.5\,kpc. The CO to SFR ratio (bottom panel) gives the dependence of the depletion timescale with the radius. The dashed line represents the ratio of the two exponential disks. The depletion time decreases slowly, a factor $\sim$2 over 4\,kpc. 

The cloud to cloud variation in the \Ha\ to CO ratio is considerably higher, about a factor 5. Molecular clouds at R$_{\mathrm{gal}}\approx$ 2.5\,kpc have a longer gas depletion timescale than clouds at R$_{\mathrm{gal}} \approx $ 3.5\,kpc (the NGC 604 region). In our map, these radii correspond to the beginning and end of the main arm, a gas-rich stellar arm ending in the giant HII region NGC\,604. To the West, the "beginning" of the main arm coincides with the "arm" just further out. It is appealing to interpret the variations in the \Ha\ to CO ratio in terms of stages in the process of molecular cloud formation, star formation and subsequent cloud destruction and to link these to flow into or out of a spiral arm.  However, the situation here is clearly too complex  and the effect  of stellar spiral arms quite limited due to their poor definition and amplitude (M\,33 being a flocculent spiral according to \citet{Elmegreen03}). \citet{Wong02} find a molecular gas depletion time scale similar to \citet{Kennicutt98} and \citet{Murgia02}, 1 -- 3\,Gyr, with radial gradients of a factor $\la 2 - 3$, like the CO/\Ha\ gradient we find.

Is the increase in the \Ha\ to CO ratio (SFE, the inverse of the depletion timescale) due to our use of a constant $\ratioo$ factor? Or is it simply that the $\ratioo$ ratio increases with radius rather than invoke a higher efficiency? Let us assume that the $\ratioo$ factor is (inversely) linear with metallicity. This is probably a reasonable assumption given that the radiation field also decreases radially.  As described in Sect 1., the metallicity gradient in M\,33 and perhaps in spiral galaxies in general is currently subject to debate.  Assuming a moderate gradient of -0.05\,dex per kpc, this increases the 1.4\,kpc CO scale length to a molecular gas scale length of nearly 1.7\,kpc, roughly the same as that of the \Ha\ emission. Using a CO scalelength of 1.7\,kpc, the ratio of the fits to the $\Sigma_{\mathrm{CO}}$ and the $\Sigma_{\mathrm{SFR}}$ is constant to within the uncertainties. The CO(2--1)/CO(1--0) line ratio may decrease with radius and our assumption of a constant line ratio would result in a slight ($\lesssim$10\%) underestimate of the H$_{2}$ scalelength. Extrapolating this scale length to the entire disk of M\,33 leads to a molecular gas mass of $\sim$1.6\dix{8}\,\msol. This is substantially higher than the \citet{Engargiola03} estimate but slightly lower than \citet{Heyer04} -- we believe that their use of a masked moment map led to an overestimate of the CO scale length.

Within the region we observed, applying a $\ratioo$ ratio varying inversely with the estimated metallicity gradient, $$ \ratiob = \frac{2}{0.75} \times 10^{20} \times 10^{0.05 \, (r-1.5(\mathrm{kpc}))} $$ and assuming our nominal conversion factor to be appropriate at 1.5\,kpc, we obtain a molecular gas mass (including He) of 2.0\dix{7}\,\msol, 25\% higher than the 1.6\dix{7}\,\msol\ found with a constant ratio. If we assume that the "standard" ratio we adopted applies at the center (i.e. $0.05 r$ in the exponent above rather than 0.05 ($r$--1.5\,kpc)), then we obtain 2.4\dix{7}\,\msol. The molecular gas depletion timescale for our region is then 2\dix{8}\,yr $\la$ M$_{\mathrm{mol}}$/SFR $\la$ 5\dix{8}\,yr, several times less than what is found for large spirals.

While the uncertainties in the SFR and the H$_2$ mass can just bring M\,33 into the range of "normal" spirals, we feel it is more likely that in fact the SFE is higher in M\,33 than in larger spirals, increasing the value of M\,33 as a potential surrogate for intermediate-redshift spirals. Sensitive CO observations are necessary of other nearby small spirals, NGC\,2403 and NGC\,6822 being the obvious candidates, to determine whether the high SFE is a general feature of small spirals. 
 
\section{Formation of Molecular Gas}

\begin{figure*}[!ht]
 \centering
 \includegraphics[width=300pt,angle=270.]{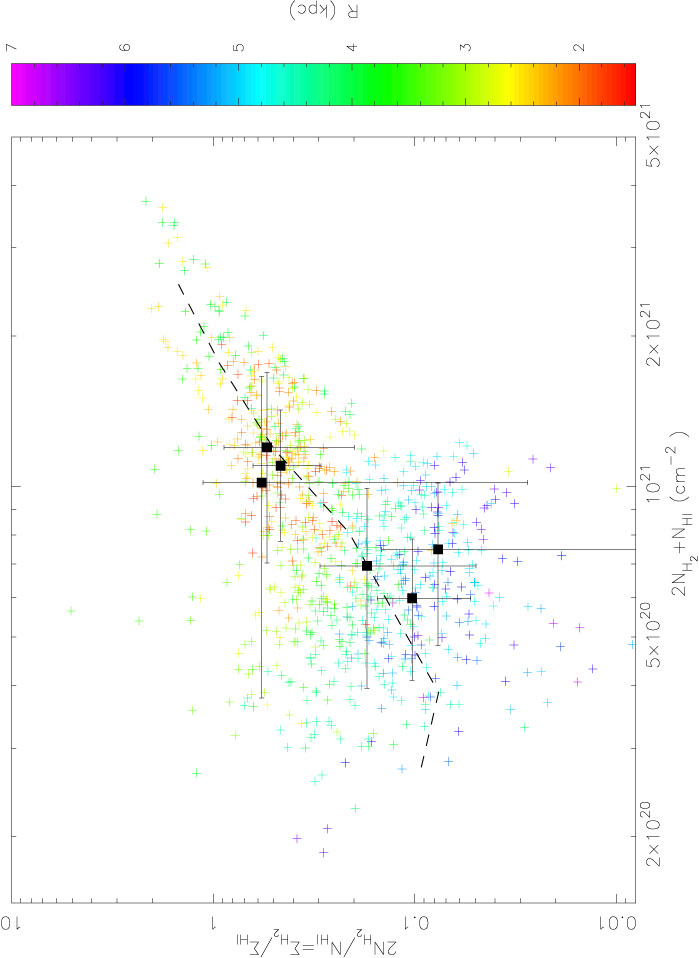}     
 \caption{Molecular fraction f$_{\mathrm{mol}}$ as a function of the deprojected total number of Hydrogen atoms N$_{H}$. Each point represents the mean value of 16\arcsec\ pixels from the 30\arcsec\ resolution map. Only pixels where CO emission is present in the 3$\sigma$ threshold map are plotted but all pixels (i.e. the unmasked map) are used to take the averages. The color bar represents the radius (R$_{\mathrm{gal}}$) of each pixel. The dashed line is the mean value of the molecular fraction per bin of N$_{H}$. Squares are the mean value of the molecular fraction per radius.}
 \label{fh2}
\end{figure*}

Figure~\ref{fh2} shows the molecular fraction f$_{\mathrm{mol}}$=2N$_{H_{2}}$/N$_{HI} = \Sigma_{\mathrm{H}_{2}}/\Sigma_{\mathrm{HI}}$ versus the total column density of Hydrogen atoms N$_{\mathrm{H}}$=2N$_{\mathrm{H}_{2}}$+N$_{\mathrm{HI}}$. We take N$_{\mathrm{H}_{2}}$ values from the 30\arcsec\ resolution CO map with a pixel size of 16\arcsec, and N$_{\mathrm{HI}}$ values in a map with the same pixel properties. We plot only points where molecular emission is expected from the 3$\sigma$ threshold map. The dashed line is the average of all pixels (without the 3$\sigma$ mask) in the map per bin of total H column density. It is thus not biased by any masking or by noise. Squares are the average values of the molecular fraction and H column density for 1\,kpc wide bins centered on 1.5, 2.5, .. 6.5 kpc. In the inner disk (R$_{\mathrm{gal}}\le$4\,kpc) the molecular fraction is similar f$_{\mathrm{H}}\simeq$0.5. At larger radii, the molecular fraction decreases regularly with radius even at roughly constant column density. The same effect is observed in the non-masked CO data set in Fig.~\ref{HI_H2} -- the ratio is constant to about 3.8\,kpc and then the H$_2$ fraction decreases regularly. The difference in the $\Sigma({\mathrm H_2})/\Sigma({\mathrm{HI}})$ values between the figures is due to the inclusion of all the CO non-detections in Fig.~\ref{HI_H2}.  

\begin{figure}[!b]
 \centering
 \includegraphics[width=180pt,angle=270.]{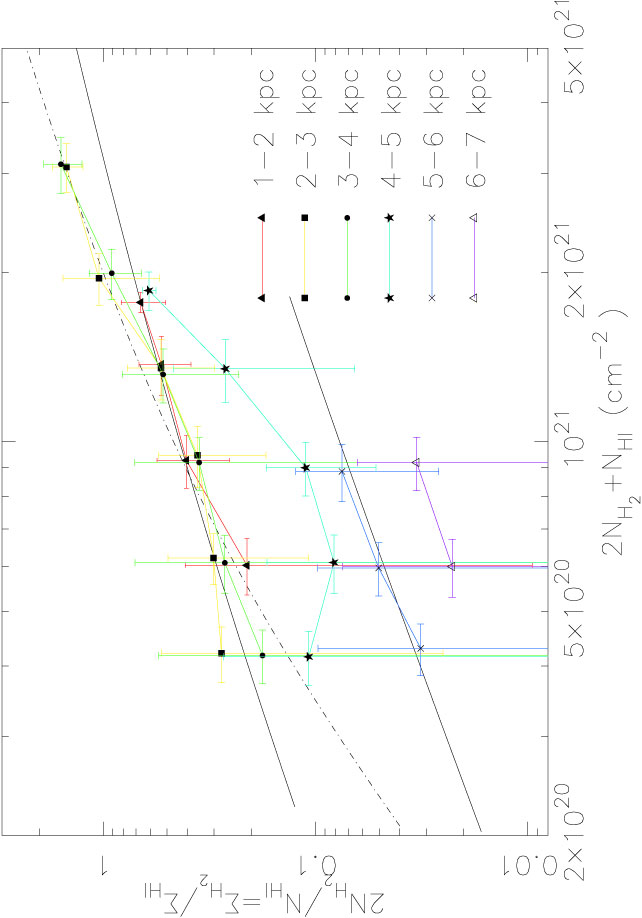}     
 \caption{Mean value of the molecular fraction averaged as a function of deprojected N$_{\mathrm{H}_{2}}$ and galactocentric radius.
Solid lines represent the power law $(2N_{H_{2}})\propto (N_{HI})^{2}$ for 2 normalisation value. The dot-dashed line is the power law
with a power index of 3. A linear relationship, 2N$_{\mathrm{H}_{2}}\propto $ N$_{\mathrm{HI}}$, would appear as a horizontal line.}
 \label{fh1}
\end{figure}

In Figure~\ref{fh1} this effect is seen even more clearly. The color coding of the points indicate the radial bins (as before) over which the molecular fraction and H-atom column density are averaged. The inner three radial bins show similar molecular fractions at similar H column densities, with the molecular fraction increasing with the total Hydrogen column density.  The more distant radial bins (4 -- 7\,kpc) show that for similar H atom column densities, the molecular fraction decreases with radius.  At all radii, the molecular fraction increases with H column density. Two black lines are shown in Fig.~\ref{fh1}; they indicate the locus of points where 2N$_{\mathrm{H}_{2}}\propto $ N$_{\mathrm{HI}}^{2}$ normalized to f$_{\mathrm{mol}}$ = 0.04 and 0.3 at H column densities of 6\dix{20}\,cm$^{-2}$. The relationship 2N$_{\mathrm{H}_{2}}\propto $ N$_{\mathrm{HI}}$ would appear as a horizontal line in Fig. ~\ref{fh1} -- this is clearly not the case. Even if the $\ratioo$ ratio increases with radius by a factor 2-3 (constant SFE), the molecular fraction still decreases significantly with radius. 
{\it Two important factors influence the formation of molecular gas: (i) the Hydrogen column density, generally dominated by the HI column density, and (ii) the galactocentric radius.} For the latter, it is not clear whether the principal physical factor is the metallicity, the radiation field, the pressure, or some other factor probably related to the above.

\section{What determines the atomic-to-molecular gas ratio?}

In addition to the Toomre-Kennicutt-Schmidt \citep{Toomre64, Kennicutt89,Kennicutt98, Schmidt59} relationship between total gas density and star formation rate, it has been suggested that the pressure is what governs whether stars form or not \citep{Elmegreen94, Wong02, Blitz06}.  In the following, we compare the pressure-based prescription as proposed by \citet{Blitz06} with our observations. In their work, the hydrostatic pressure controls whether atomic gas is converted into molecular gas, which then rapidly forms stars. The hydrostatic pressure is \citep[Equation 1,][]{Blitz06}
\begin{center}
\begin{eqnarray}
  \label{pressure}
  P = (2G)^{0.5}\Sigma_{g}v_{g}\big[\rho_{\star}^{0.5}+\big(\frac{\pi}{4}\rho_{g}\big)^{0.5}\big]
\end{eqnarray}
\end{center}
In which they assume $\rho_{\star}>>\rho_{g}$ and neglect the $\rho_{g}$ term. They found that for their sample of galaxies, including M\,33, this worked rather well.

\begin{figure}[!ht]
    \centering
    \includegraphics[width=250pt]{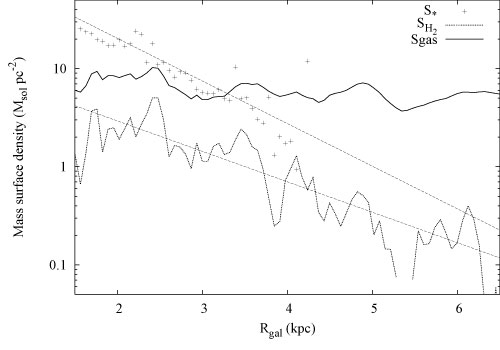}
    \caption{Azimuthal average of the deprojected mass surface density of the H$_{2}$, the gas (HI+H$_{2}$) and the stars. The K band image (2MASS) is averaged on the CO area and converted into stellar mass surface desity using the equation~\ref{K_to_M} \citep{Blitz06}. The stellar disk has a radial scale length of 1.0$\pm$0.05\,kpc.}
    \label{Kaverage}
    \end{figure}

We are interested in the outer parts of spiral galaxies where the assumption that the gas surface density is much less than the stellar surface density does not hold.  In M\,33, beyond about 3.5 kpc, the gas surface density, dominated by the HI for just about any reasonable $\ratioo$ factor, is actually greater than the stellar surface density, assuming a mass-to-light ratio in the K$'$ band of 0.5\,M$_\odot$/L$_{\odot_{,K'}}$. With the exception of the $1.5 \la r \la 2.5$\,kpc range, our data do not test the \citet{Blitz06} pressure-based molecular fraction but rather allow us to extend {\it much further} the range in radii and surface density (hence pressure) over which the molecular fraction can be measured. In Fig.~\ref{Kaverage} we average the K band emission over the CO area. Assuming a stellar mass to light ratio of 0.5, the conversion of the surface brightness to mass is \citep[e.g.][]{Blitz06}:
\begin{eqnarray}
 \label{K_to_M}
 \mathrm{log\,M}_{*} = -0.4\mu_{K}+ 9.62 + \mathrm{log\,cos}\,i 
\end{eqnarray}
where $\mu_{K}$ is the surface brightness in mag arcsec$^{-2}$ and $i$ the inclination, the stellar mass M$_{*}$ is then given  in M$_{\odot}$\,pc$^{-2}$.

Further out, where the gas mass exceeds the stellar mass, the molecular fraction decreases rapidly, varying as the CO intensity which decreases roughly exponentially (the HI remaining roughly constant). The HI surface density, and thus total gas surface density because the HI is dominant in M\,33 particularly beyond about 3.5\,kpc, is roughly constant so the pressure should not decrease by more than about a factor 2. This comes  from allowing the stellar term (first term in brackets of Equation~\ref{pressure}) to disappear.  The structure of the HI component is not sufficiently well known to say that the mid-plane pressure is not a major factor because we cannot easily separate the cold from warm HI components and determine their scale heights and velocity dispersions, both of which enter into the pressure calculation.  However, the simple formalism based on pressure assuming that the scale heights and velocity dispersions remain constant with galactocentric radius does not fit the molecular fraction in M\,33, at least in the outer parts where the gas pressure dominates.

{\it Does the metallicity gradient, and potential gradient in the $\ratioo$, affect the analysis above?} A molecular scale length of 1.7\,kpc has little effect on the total gas surface density due to the domination by the HI and as such only changes the rate at which the molecular fraction declines over a region where the pressure (dominated by the HI) is approximately constant.  

The CO to HI ratio (Fig.~\ref{HI_H2}) shows a break at 3.5 -- 4\,kpc. This value corresponds to the radius where $\Sigma_{*}$=$\Sigma_{\mathrm{gas}}$. It also corresponds to the end of identifiable stellar spiral arms. A sensitive map of the whole disk of M\,33 would be necessary to see if this "break" is a feature linked to the distance from the center (or presence of stellar arm) or if it reflects a peculiarity of the region we have covered, such as simply the border of a high gas column density region.

\section{Conclusions}

We present deep $^{12}$CO(2--1) observations of the North part of M\,33. The observations were carried out with HERA, a multibeam instrument at the IRAM 30m telescope. We observed an 700\arcsec$\times$1300\arcsec\ area of the North part of M\,33 with a resolution of 11\arcsec. The observations cover a range of galactic radii from 1.5 to 7\,kpc, reaching the R$_{25}$ radius. The high sensitivity of HERA permits us to reach a mean noise level of 12\,mK in the northern part. Since most of the observations are not sensitive enough to find molecular emission in the outer parts of galaxies, this is the first study of SFE in the outer disk of galaxies.

\begin{itemize}

\item Despite the low metallicity and the low radiation field in the outer disk, we detect molecular clouds at large galactocentric radii.
The furthest detection is at R$_{\mathrm{gal}}$=6.3\,kpc (close to the optical radius).

\item  We estimate the H$_2$ mass in our zone to be 1.6\dix{7}\,\msol, using a conversion factor of $\ratioa = $2\dix{20}\,\Xunit and assuming a line ratio of CO(2--1)/CO(1--0) = 0.75. We used the \Ha, NUV and FIR emission to derive the SFR of M\,33. In this region we find a SFR of 0.05 -- 0.1\,\mpc. The molecular gas depletion timescale is then 1.6 -- 3.2\dix{8}\,yr. {\it M\,33 is more efficient in forming stars than normal local universe spiral galaxies.}

\item The molecular emission is mostly associated with HI column density peaks. Near the center, the HI to H$_{2}$ ratio is roughly constant and increases in the outer part of the disk. The CO emission correlates well with tracers of star formation (IR, UV and \Ha). 

\item The molecular-to-atomic gas mass ratio ($f_{\mathrm{mol}}$) is roughly constant at $r \la 3.5$ kpc and then decreases sharply with galactocentric radius.  Examining the dependence of $f_{\mathrm{mol}}$ on radius and Hydrogen column density (N$_{\mathrm{H}}$), we find that at any given radius $f_{\mathrm{mol}}$ increases with N$_{\mathrm{H}}$ and that at a given N$_{\mathrm{H}}$, the molecular fraction decreases with radius in the outer parts but remains roughly constant in the inner parts.  The increase (at any given radius) in $f_{\mathrm{mol}}$ is consistent with a dependence of N(H$_2$) $\propto$ N(HI)$^2$.  If all radii are mixed then the dependence 
appears (artificially) much steeper.

\item A critical point of this study is the value of the CO-to-H$_2$ conversion factor. We consider a "standard" value of 2\dix{20}\,\Xunit, appropriate for a solar metallicity. Allowing for a metallicity gradient of -0.05\,dex\,kpc$^{-1}$ and assuming the $\ratioo$ factor to be inversely proportional to the metallicity, the radial scalelength of the H$_2$ distribution is then 1.7\,kpc.  This brings the scale length of the molecular gas distribution to roughly that of the star formation as traced by the \Ha. The total H$_2$ ($+$He) mass in the region we have observed becomes 2\dix{7}\,\msol\ and the gas depletion timescale increases by 25\%. This does not change the primary conclusion that the SFE in M\,33 is particularly high.
  
\item The azimuthal average of the CO mass surface density in the area we mapped is roughly fit by an exponential disk with a scale length of R$_{\mathrm{CO}}$=1.4$\pm$0.1\,kpc and a central mass surface density of $\Sigma_{0}$ of 10$\pm$1\,M$_{\odot}$\,pc$^{-2}$. Integration of the disk gives a total molecular mass of M$_{H_{2}}$=1.2$\pm$0.3\,\dix{8}\,\msol. However, our observations do not include the central regions (nor the southern part of the galaxy) so the extrapolation to the whole disk may not be appropriate.

\item The use of a multibeam receiver makes it feasible to map large areas in the On-The-Fly observing mode. Because we mapped both arm and inter-arm regions and not simply the regions in which one would expect to detect CO emission, we detected an inter-arm molecular cloud 
(at $\alpha$=\hms{01}{34}{16.9}, $\delta$=\dms{30}{59}{31.4}). This cloud is not associated with a peak in HI column density nor with identifiable star formation. The mass of this cloud is M$_{\mathrm{H}_{2}} \sim $ 2\dix{4}\,\msol. 

\item Is the molecular gas fraction controlled by the hydrostatic pressure? Over the region we have observed in CO, the inner part is where we can directly compare with \citet{Blitz06} because they only consider regions where the stellar surface density is much greater than the gaseous. Although the stellar mass decreases sharply from 1.5 to 4 kpc, the molecular-to-atomic gas mass ratio remains roughly constant for any given column density (although the molecular fraction increases with column density for a given radius).  One should note that we have not observed the whole disk so the above is true only if our region is representative. Further out, the molecular fraction drops with radius and does not seem to follow the pressure-based prescription.
 
\end{itemize}

\acknowledgements{We would like to thank J.M. van der Hulst for making the \citet{Deul87} HI cube available, 
Ren\'e Walterbos for providing the \Ha\ image, and M. Haas and H. Hippelein for the ISO FIR images.
We acknowledge the usage of the HyperLeda database (http://leda.univ-lyon1.fr) and thank the IRAM staff in Granada for their help with the observations.}

\bibliography{7711bib}
\bibliographystyle{aa}

\end{document}